\documentclass[a4paper,12pt]{article} 
\usepackage{amssymb,amsmath}

\newcommand{\RR}{{\mathbb{R}}}
\newcommand{\HH}{{\mathbb{H}}}
\newcommand{\pa}{\partial}
\newcommand{\half}{\tfrac{1}{2}}
\newcommand{\diag}{\mathop{\mathrm{diag}}\nolimits}

\title{Infinite-Parameter ADHM Transform}
\author{R.\ S.\ Ward\footnote{Email: richard.ward@durham.ac.uk}
  \bigskip
  \\Department of Mathematical Sciences,
  \\Durham University, Durham DH1 3LE.}
\date{\today}

\begin{document}

\maketitle

\begin{abstract}
\noindent
 The Atiyah-Drinfeld-Hitchin-Manin (ADHM) transform and its various
 generalizations are examples of non-linear integral transforms between
 finite-dimensional moduli spaces. This note describes a natural
infinite-dimansional generalization, where the transform becomes a map
from boundary data to a family of solutions of the self-duality equations in
a domain.
\end{abstract}

\begin{center}
  {\sl Dedicated to Michael Atiyah, in Memoriam}
\end{center}
  
\section{Introduction}

One of many discoveries named after Michael Atiyah is the ADHM
(Atiyah-Drinfeld-Hitchin-Manin) transform \cite{ADHM78}. Starting with
the work of Nahm \cite{N80, N83}, it was subsequently generalized
in various ways; see for example the review \cite{J04}. Independent of
this, but related to it, was the Fourier-Mukai transform in algebraic
geometry \cite{Muk81, BBH09}.

Now ADHM is an integral transform, and as such is analogous
to the Fourier, Radon and Penrose transforms, and also to the inverse
scattering transform in soliton theory. These other examples are usually
encountered in infinite-dimensional contexts: for example, the Fourier
transform is an isomorphism of infinite-dimensional vector spaces.
By contrast, the ADHM transform and its generalizations
have mostly been studied in finite-dimensional contexts, giving
correspondences between finite-dimensional moduli spaces. Indeed, it
originally arose from algebraic geometry (the construction of algebraic
vector bundles), and index theory was an important part of it. The ADHM
construction and its various generalizations involve the kernels of Dirac
operators, with part of the analysis being to prove that these operators are
Fredholm, so that their kernels are finite-dimensional.

The purpose of this note is to point out that ADHM can in fact operate
comfortably as an infinite-dimensional transform, just like its cousins.
In fact, early work \cite{Wit79, N80, Os82} already suggested an
underlying local and infinite-dimensional structure. Of course, it has
always been clear that one may take the basic ADHM transform, in which ADHM
data of rank $N$ correspond to instantons of charge $N$, and then
implement some sort of $N\to\infty$ limit to obtain
an infinite-dimensional system; but the aim here is to go beyond this
bald observation, and to describe a specific scheme.


\section{The Original ADHM Transform}

This section summarizes the original ADHM construction \cite{ADHM78},
in order to describe the background and establish notation. Let
$x_\mu=(x_1,x_2,x_3,x_4)$ denote Cartesian coordinates on $\RR^4$,
and let $A_\mu$ denote a gauge potential.
For simplicity, we take the gauge group to be SU(2); so each of
$A_1$, $A_2$, $A_3$ and $A_4$
is a $2\times2$ antihermitian tracefree matrix,
and $A_\mu$ describes an SU(2) connection over $\RR^4$.
The corresponding gauge field (curvature) is
\[
  F_{\mu\nu}=\pa_\mu A_\nu - \pa_\nu A_\mu + [A_\mu , A_\nu],
\]
where $\pa_\mu=\pa/\pa x_\mu$.
The self-dual Yang-Mills equation is the condition that this 2-form
be self-dual on Euclidean $\RR^4$, namely that
\begin{equation} \label{SDYM}
    \half \varepsilon_{\alpha\beta\mu\nu}F_{\alpha\beta} = F_{\mu\nu}.
\end{equation}
Here we are employing the Einstein convention of summing over repeated indices,
and $\varepsilon$ denotes the standard totally-skew tensor.
The self-duality equations~(\ref{SDYM})
constitute a set of nonlinear partial differential equations for $A_\mu$,
invariant under gauge transformations
 $A_\mu \mapsto \Lambda^{-1}A_\mu\Lambda+ \Lambda^{-1}\pa_\mu\Lambda$
with $\Lambda:\RR^4\to{\rm SU(2)}$, and also
conformally invariant. If we impose boundary conditions which amount to saying
that the field extends to the conformal compactification $S^4$, then
the solutions are called instantons, and are classified topologically
by an integer~$N$.

There is an \lq ansatz\rq\ (the 't Hooft-Corrigan-Fairlie-Wilczek ansatz \cite{CF77})
which produces a subset of the solutions of~(\ref{SDYM}). It has the
form
\begin{equation} \label{ansatz}
   A_\mu = \half T_{\mu\nu} \, \pa_\nu \log\phi,
\end{equation}
where $T_{\mu\nu}$ (defined below) is a certain constant tensor with
values in the Lie algebra $\mathfrak{su}(2)$. Assuming this form for $A_\mu$
reduces~(\ref{SDYM}) to the Laplace equation
\begin{equation} \label{Laplace}
  \pa_\mu \pa_\mu \phi = 0
\end{equation}
on $\RR^4$. This is a local result: no boundary conditions are required.
To get an $N$-instanton solution, one may take $\phi$ to be a sum of $N$
fundamental solutions
\begin{equation} \label{phi}
   \phi(x) = 1 + \sum_{a=1}^{N} \frac{\lambda_a^2}{|x-x^{(a)}|^2},
\end{equation}
where the $x^{(a)}$ denote $N$ distinct points in $\RR^4$,
$|x|^2=x_\mu x_\mu$ denotes the Euclidean length-squared, and the
$\lambda_a$ are positive weights. The singularities at $x=x^{(a)}$ are
removable by a gauge transformation on $A_\mu$. This simple
formula gives a $(5N)$-parameter family of instanton solutions,
which we may think of as $N$ instantons with locations $x^{(a)}\in\RR^4$
and sizes $\lambda_a$, all in phase with one another.
In the general $N$-instanton solution, which is not of the special
form~(\ref{ansatz}), each instanton acquires an individual SU(2)
phase: this gives an extra $3N$ parameters, so the full moduli space
has dimension $8N$ (or $8N-3$, if we remove an overall phase).

The relevant aspects of the ADHM construction of SU(2) instantons may
be summarized as follows \cite{ADHM78, CSW78, CFGT78, CG84}; we
mostly adhere here to the conventions of \cite{CG84}.
It is convenient to use quaternions. Let $e_a=(e_1,e_2,e_3)$ denote
quaternions with $e_1 e_2=-e_3=-e_2 e_1$, $(e_1)^2=-1$ etc, and define
$e_\mu=(e_1,e_2,e_3,1)$. A point $x\in\RR^4$
corresponds to the quaternion $x = x_\mu e_\mu \in \HH$.
The quaternion-valued 2-form $T_{\mu\nu}$ is defined by
$e^*_\mu e_\nu = \delta_{\mu\nu} + T_{\mu\nu}$, where the \lq star\rq\
denotes quaternionic conjugation (and will denote conjugate
transpose when applied to a matrix). We identify the imaginary quaternions
with the Lie algebra $\mathfrak{su}(2)$; in terms of the Pauli matrices
$\sigma_a$, one may use the identification $e_a\equiv i\sigma_a$.
So $T_{\mu\nu}$ is also a 2-form with values in $\mathfrak{su}(2)$, and
this is the object appearing in~(\ref{ansatz}).

Now let $L$ be an $N$-row vector of quaternions, and $M$
a symmetric $N\times N$ matrix of quaternions. They are required to 
satisfy the ADHM constraint, namely that
\begin{equation} \label{ADHMeqn}
   L^* L + M^* M \mbox{ is real.}
\end{equation}
We also need an invertibility condition, to get a non-singular gauge field;
but~(\ref{ADHMeqn}) is the crucial condition for generating solutions
of~(\ref{SDYM}).
The ADHM data for an $N$-instanton solution consist of $(L,M)$,
subject to~(\ref{ADHMeqn}), and modulo the equivalence
\begin{equation} \label{equiv}
  M \equiv R^t M R, \quad L \equiv \alpha L R,
\end{equation}
where $R\in{\rm O}(N)$ and $\alpha$ is a unit quaternion.

To obtain an instanton gauge potential $A_\mu$ from these data, we proceed
as follows. For each $x\in\HH$, let $v$ be an $N$-row vector of quaternions,
depending on $x$, and satisfying the linear equation
\begin{equation} \label{v-eqn}
   L + v (M+x) = 0.
\end{equation}
Then set
\begin{equation} \label{A}
   A_\mu = \frac{1}{2\phi}\left[v\,(\pa_\mu v^*) - (\pa_\mu v)\,v^*\right],
\end{equation}
where $\phi=1+vv^*$. This field $A_\mu$ is a 1-form with values in the
imaginary quaternions, and hence in $\mathfrak{su}(2)$; and it is an instanton
gauge potential. This is the ADHM transform: it in fact gives a one-to-one
correspondence between ADHM data $(L,M)$ of rank $N$, and SU(2)
$N$-instantons up to gauge equivalence.

To get the special class of solutions~(\ref{ansatz}), we take
the components $\lambda_a$ of $L$ to be real and positive, and
$M=\diag(-x^{(1)},-x^{(2)},\ldots,-x^{(N)})$; these data satisfy the
ADHM constraints~(\ref{ADHMeqn}).
The solution of~(\ref{v-eqn}) is then $v=[v_1,\ldots,v_N]$, where
\[
  v_a=\lambda_a(x^{(a)}-x)^{-1};
\]
and evaluating~(\ref{A}) gives~(\ref{ansatz}) and~(\ref{phi}).


\section{Generalized ADHM --- Ansatz Case}

Our aim is to describe an infinite-parameter generalization of the
standard ADHM transform.
The starting-point is the ansatz~(\ref{ansatz}), which produces a
subclass of self-dual gauge fields from harmonic functions $\phi$. The ADHM
construction involves harmonic functions of the form~(\ref{phi}), namely
finite sums of fundamental solutions. By contrast, the general smooth solution
of~(\ref{Laplace}) in a domain $D$ in $\RR^4$ is determined by arbitrary
functions, namely data on the boundary $S=\partial D$.
This suggests a generalization of ADHM in which the data, or at least
most of it, correspond to arbitrary functions on the 3-surface~$S$. 
It is this idea that we shall pursue in what follows, beginning with the
ansatz~(\ref{ansatz}), and then moving on to a more general class of fields.
For simplicity we will take $D$ to be the unit ball $|x|\leq1$, so that $S$ is
the unit 3-sphere $|x|=1$; however, the idea works just as well for other
domains.

The self-duality equation~(\ref{SDYM}) can be re-written as a non-linear
second-order equation for a Hermitian matrix-valued function on $D$;
this form is sometimes referred to as the Hermitian Yang-Mills equation.
It is known that the corresponding Dirichlet problem has a unique solution
\cite{D92}. So for example in the SU(2) case, the general (local) solution
of~(\ref{SDYM}) depends on three arbitrary functions of three variables,
namely the boundary data.
The aim in this note is to describe a generalized ADHM transform which
may be viewed as an implementation of the map from boundary data
to the solution in the interior.

Our generalized setup is as follows. As coordinates on $S$ we use unit
quaternions: $y\in\HH$ with $|y|=1$. The $N$-vector $v$ of the previous
section becomes a square-integrable quaternion-valued function $v(y)$
on $S$, depending also on $x$. The real $N$-vector $L$ becomes a
smooth real-valued function $L(y)$ on $S$. The linear
equation~(\ref{v-eqn}) defining $v$ is replaced by
\begin{equation} \label{v-inf-ansatz-eqn}
   L(y) + v(y) (-y+x) = 0.
\end{equation}
So, in effect, the matrix $M$ has become $-y$ times a three-dimensional
delta-function on $S$. Clearly the solution of~(\ref{v-inf-ansatz-eqn}) is
\begin{equation} \label{v-inf-ansatz-soln}
  v(y)=L(y)(y-x)^{-1}.
\end{equation}
To get a gauge potential $A_\mu(x)$ from $v$, we
define a quaternionic product by
\[
  \langle v, w\rangle = \int_S v(y)\, w(y)^* \, |dy|^3,
\]
where $|dy|^3$ denotes the standard Euclidean measure on $S=S^3$;
and then set
\begin{equation} \label{A-inf}
 A_\mu=\frac{1}{2\phi}\left[\langle v,\pa_\mu v\rangle
                  -\langle\pa_\mu v,v\rangle\right],
\end{equation}
where $\phi$ is the real-valued function $\phi(x)=1+\langle v,v\rangle$. As before,
the partial derivatives $\pa_\mu$ in~(\ref{A-inf}) are with respect to $x_\mu$.
With $v$ given by~(\ref{v-inf-ansatz-soln}), the function $\phi$ is
\begin{equation} \label{phiR3}
  \phi(x) = 1 +
          \int_S \frac{L(y)^2}{|x-y|^2} \, |dy|^3,
\end{equation}
and the expression~(\ref{A-inf}) then reduces to the ansatz
form~(\ref{ansatz}), with $\phi$ given by~(\ref{phiR3}).

Now~(\ref{phiR3}) is simply the Green's function
formula for solutions of the Laplace equation in the domain $D$,
with Robin boundary data on $S=\pa D$. More precisely, $L(y)$
is determined by $\phi$ as
\begin{equation} \label{Green}
  2\pi^2 L(y)^2=\phi(y)+\pa_n\phi(y)-1,
\end{equation}
where $\pa_n\phi=y_\mu\partial_\mu\phi$ denotes the outward normal
derivative of $\phi$ on $S$. So for this class of self-dual gauge fields,
the ADHM data, consisting essentially of the real-valued function $L$
up to sign, may be interpreted as boundary data for the gauge field.

One may make contact with the original, finite, version~(\ref{phi}) as follows.
Suppose that we are given a real-valued function $L$ on $S$, giving rise to the
harmonic function $\phi$ as in~(\ref{phiR3}), and the  gauge potential $A_\mu$.
For each positive integer $N$, choose a uniformly-distributed sample of $N$ points
$x^{(1)},\ldots,x^{(N)}$ on $S$, and define \lq finite\rq\ ADHM data
$L^{(N)}=[\lambda_1,\ldots,\lambda_N]$ and $M^{(N)}$ by
\begin{equation} \label{LM_MC}
  \lambda_i=\pi\sqrt{\frac{2}{N}}\,L(x^{(i)}), \quad
  M^{(N)}=\diag(-x^{(1)},-x^{(2)},\ldots,-x^{(N)}).
\end{equation}
Then the standard ADHM construction gives an $N$-instanton
field $A_\mu^{(N)}$ as in (\ref{ansatz}) and (\ref{phi}), involving the
harmonic function
\begin{equation} \label{phiMC}
   \phi^{(N)}(x) = 1 + \sum_{i=1}^{N} \frac{\lambda_i^2}{|x-x^{(i)}|^2}.
\end{equation}
Now $\phi^{(N)}\to\phi$ as $N\to\infty$, this being simply a Monte Carlo
evaluation of the integral (\ref{phiR3}); and therefore we also have
$A_\mu^{(N)}\to A_\mu$ as $N\to\infty$. In effect, the expression (\ref{phiMC}) 
approximates the general ansatz solution in $D$ in terms of $N$ instantons
on the boundary $S$, all in phase with one another.

This picture could be extended further: in the Green's formula,
we could add a finite number of delta-function sources in the interior of $D$.
Then $\phi$ would have singularities of the form~(\ref{phi}) inside $D$, but
would still produce a smooth gauge field, incorporating instantons inside $D$.
We then get a combination of the
\lq finite-parameter\rq\ original version of the ADHM construction
and the \lq infinite-parameter\rq\ version introduced above. 
This extended story is reminiscent of the inverse scattering transform
for solitons, where the scattering data consists of a finite-parameter soliton
part, plus an infinite-parameter radiation part. In our case, we would have
a finite number of instantons located inside $D$, plus infinitely many other
degrees of freedom.


\section{Generalized ADHM --- Full Version}

The aim here is to extend the structure of the previous section, so that
it is no longer restricted to fields of the ansatz type~(\ref{ansatz}).
By analogy with the standard case, this can be done by allowing $L(y)$ to
be quaternion-valued rather than real-valued. Then $M$ will no longer
be \lq diagonal\rq, but rather becomes a symmetric quaternion-valued generalized
function $M(z,y)$, thought of as the kernel of an integral operator acting
on the function $v$ by
\begin{equation} \label{vM}
  (vM)(y) = \int_S v(z) M(z,y) \, |dz|^3.
\end{equation}
The data $(L,M)$ are required to satisfy the analogue of~(\ref{ADHMeqn}),
namely that
\begin{equation} \label{ADHMeqn-inf}
  L(y)^*L(z) + \int_S M(y,s)^* M(s,z) \, |ds|^3 \mbox{ is real}.
\end{equation}
This will guarantee that the corresponding gauge field is self-dual,
wherever it is defined. The linear equation satisfied by $v$ then becomes
\begin{equation} \label{v-eqn-inf}
   L + v M + vx = 0,
\end{equation}
and the self-dual gauge potential is given, as before, by~(\ref{A-inf}).

It is straightforward to generalize the \lq discrete approximation\rq\
(\ref{LM_MC}, \ref{phiMC}) by taking a sample of $N$ points on $S$,
and this illustrates how the structure described above may be viewed as
an $N\to\infty$ limit of the $N$-instanton construction. 

Now in the general case, we have the problem of solving the non-linear
ADHM constraint equation~(\ref{ADHMeqn-inf}), and this is difficult: indeed,
already very difficult for the original finite system~(\ref{ADHMeqn}).
In the discussion below, we first examine the simplest example; and then
describe the linearized version, which gives an indication of \lq how many\rq\
self-dual gauge fields the construction produces.

The simplest non-trivial example with $L(y)$ not real-valued is
\begin{equation} \label{simpleLM}
   L(y)=\kappa y, \quad 
    (vM)(y)=-v(y)y+\lambda \int_S v(z)(y+z) \, |dz|^3,
\end{equation}
where $\kappa$ and $\lambda$ are real constants satisfying
\[
   2\pi^2\lambda^2-2\lambda+\kappa^2=0, \quad
    0<\kappa\pi\sqrt{2}<1, \quad 0<2\pi^2\lambda<1.
\]
It is easily checked that this $(L,M)$ satisfies the constraint
equation~(\ref{ADHMeqn-inf}).
Clearly there is a high degree of symmetry in this solution, in particular under
the rotation group SO(4) acting on $S$; and this implies that
the corresponding gauge field must be the 1-instanton located at
the origin $x=0$. The only parameter in this solution is the instanton size,
and this is determined by $\kappa$, or equivalently by $\lambda$. Explicitly
implementing the ADHM construction with $x=0$, to obtain the gauge field
there, reveals that the instanton size is in fact given by
\[
  \rho=\frac{1-2\pi^2\lambda}{\sqrt{2\lambda(1-\pi^2\lambda)}}
\]
in terms of $\lambda$. Note that $\rho\to0$ as $\lambda\to1/(2\pi^2)$,
and $\rho\to\infty$ as $\lambda\to0$.
As was pointed out previously, one can obtain the 1-instanton
at $x=0$ by putting it in \lq by hand\rq\ as a delta-function source.
But we see here that solutions including instantons inside $D$ can
also, and perhaps more neatly, be obtained from smooth
data such as~(\ref{simpleLM}).

Now let us consider the linearized version. The details of this are somewhat
analogous to those of the finite (instanton) case described in \cite{CSW78}.
Let $\epsilon$ be a parameter with $0<\epsilon\ll1$, and take $L$ to be a
quaternion-valued function with $|L(y)|=O(\epsilon)$: we shall work to
lowest order in $\epsilon$. (The scale is set by the volume of $S$, which here
is of order unity.) If we write
\[
  M(y,z)=-y\,\delta(y-z) + P(y,z),
\]
then the constraint equation~(\ref{ADHMeqn-inf}) is equivalent to
\begin{eqnarray*}
  2(y^*-z^*)P(y,z) &=& L(y)^*L(z)-L(z)^*L(y) \\
  & & {} +\int_S\left(P(y,s)^*P(z,s)-P(z,s)^*P(y,s)\right)\,|ds|^3,
\end{eqnarray*}
and this can be solved iteratively for $P$, order-by-order in $\epsilon$.
For our purposes here, it is sufficient to observe that $P(y,z)=O(\epsilon^2)$.
The solution of~(\ref{v-eqn-inf}) then has the form
\[
  v(y)=L(y)(y-x)^{-1} +O(\epsilon^3),
\]
and this can then be used in~(\ref{A-inf}) to compute the leading term 
${\cal A}_\mu$ in the
gauge potential $A_\mu$, which will be of order $\epsilon^2$. The calculation
is straightforward, and the details will be omitted here; the result can be
written as follows.

Define $C_\mu^a[\phi]$ by
\[
   C_\mu^a[\phi]e_a = \half T_{\mu\nu}\pa_\nu\phi,
\]
which is just the linearized version of~(\ref{ansatz}). So $C_\mu^1$
produces a self-dual U(1) gauge field $a_\mu=C_\mu^1[\phi]$ from a
harmonic function $\phi$; and furthermore,
given any self-dual U(1) gauge field $a_\mu$, there exists a
harmonic function $\phi$ such that $a_\mu=C_\mu^1[\phi]$.
The same is true of $C_\mu^2$ and $C_\mu^3$. Now given $L(y)$,
define a quaternion-valued solution $\Phi$ of the Laplace equation
(\ref{Laplace}) by
\begin{equation} \label{Phi}
  \Phi(x) =1+\int_{\RR^3} \frac{L(y)^2}{|x-y|^2} \, |dy|^3,
\end{equation}
and let $\Phi_\mu$ be its quaternionic components, in other words
$\Phi=\Phi_\mu e_\mu$. Then our SU(2) gauge potential is
$A_\mu={\cal A}_\mu^a e_a+O(\epsilon^4)$, where
\begin{equation} \label{A-gen-lin}
  {\cal A}_\mu^a = C_\mu^a[\Phi_4]+\varepsilon_{abc}C_\mu^b[\Phi_c].
\end{equation}
In other words, the leading $O(\epsilon^2)$ part of $A_\mu$ just consists
of the three self-dual U(1) gauge fields ${\cal A}_\mu^a$ given by the
formula~(\ref{A-gen-lin}). Note that the generalized ADHM data $L(y)$
corresponds to boundary data, via the obvious quaternionic
generalization of the formula~(\ref{Green}).

Now if $L$ is real-valued, then~(\ref{Phi}) becomes~(\ref{phiR3}),
and~(\ref{A-gen-lin}) just reduces to the linearized version of~(\ref{ansatz}),
as expected. In this case, there is effectively only one
independent gauge field: for example ${\cal A}_\mu^1$ determines
${\cal A}_\mu^2$ and ${\cal A}_\mu^3$. To put this another way,
each of the three ${\cal A}_\mu^a$ is determined by the single harmonic
function $\Phi_4$.

If $L$ is quaternion-valued rather than real, then there are four harmonic
functions $\Phi_\mu$ appearing in the formula~(\ref{A-gen-lin}). So one
might expect to obtain a more general class of linearized self-dual SU(2) fields
than those corresponding to the ansatz~(\ref{ansatz}). This is indeed the
case, but only partially: two of the resulting ${\cal A}_\mu^a$ are independent,
but not all three of them. To see that at least two of the ${\cal A}_\mu^a$
are independent is easy: for example, given ${\cal A}_\mu^1$ and
${\cal A}_\mu^2$, set $\Phi_2=\Phi_3=0$, choose $\Phi_4$ such that 
${\cal A}_\mu^1 = C_\mu^1[\Phi_4]$, and then choose $\Phi_1$ such
that ${\cal A}_\mu^2 = C_\mu^2[\Phi_4]+C_\mu^3[\Phi_1]$.
In other words, the formula~(\ref{A-gen-lin}) can produce arbitrary
${\cal A}_\mu^a$ for $a=1,2$. But it cannot do so for $a=1,2,3$;
and consequently~(\ref{A-gen-lin}) does not yield the most general
linearized self-dual SU(2) fields in $D$. This is somewhat less obvious;
a sketch of the reasoning is as follows. If one could generate three
independent self-dual U(1) gauge fields ${\cal A}_\mu^a$ via
(\ref{A-gen-lin}), then in particular one could get
${\cal A}_\mu^1={\cal A}_\mu^2=0$, while ${\cal A}_\mu^3\neq0$.
But imposing ${\cal A}_\mu^1={\cal A}_\mu^2=0$ in (\ref{A-gen-lin})
leads, after some algebra, to ${\cal A}_\mu^3=0$.

It is reasonable, therefore, to conjecture that the analogous result is true for
the full nonlinear system, namely that the generalized ADHM construction
described in this section produces an infinite-dimensional class of solutions
of~(\ref{SDYM}) in $D$ from their boundary data: a class larger than
that of the ansatz~(\ref{ansatz}), but not the whole solution space.
In fact, the conjecture is that we get a family of solutions depending on two arbitrary
functions of three variables, whereas the ansatz solutions depend on one
such function, and the general self-dual gauge field on three such functions.


\section{Comments}

The aim above has been to describe a specific infinite-dimensional version
of the ADHM construction, and to promote the claim that Atiyah's
original finite-dimensional algebraic-geometrical picture extends
rather naturally to a local infinite-dimensional one.
This generalization could, in some sense, be viewed as
a non-linear version of the Green's function formula for solutions of the
Laplace equation on a bounded domain $D$ in $\RR^4$, in terms of
boundary data on $\pa D$. The scheme described above is not a
complete solution of the boundary-value problem: it does not yield the
general solution in the interior. So an intriguing question is whether there
is a further generalization of ADHM which does amount to a complete solution.

Another local and infinite-dimensional generalization of the
ADHM construction, which on the face of it is quite distinct from
the one presented
here, involves the Nahm equations with values in the Lie algebra
$\text{sdiff}(S^2)$ of Hamiltonian vector fields on $S^2$, and their
relation to \lq abelian monopole bags\rq\ 
\cite{W90, D10, H12, HPS12, BHS15}. In this case, we have a hodograph
transformation which transforms the Nahm equation to the Laplace equation on
a domain in $\RR^3$ (or in 3-dimensional hyperbolic space). Although
this looks rather different, it may possibly be related to
the four-dimensional scheme described above.



\begin{thebibliography}{99}

\bibitem{ADHM78}
M F Atiyah, V G Drinfeld, N~J~Hitchin and Y I Manin, Construction of
instantons. {\it Phys Lett A} {\bf65} (1978) 185--187.

\bibitem{BBH09}
C Bartocci, U Bruzzo and D Hern\'andez Ruip\'erez,
 {\it Fourier-Mukai and Nahm Transforms in Geometry and Mathematical Physics.}
 (Birkh\"auser, Boston, 2009)
 
\bibitem{BHS15} S Bolognesi, D G Harland and P M Sutcliffe, Magnetic
  bags in hyperbolic space. {\it Phys Rev D} {\bf92} (2015) 025052.

\bibitem{CSW78}
N H Christ, N K Stanton and E J Weinberg, General self-dual Yang-Mills
  solutions. {\it Phys Rev D} {\bf18} (1978) 2013--2025.

\bibitem{CFGT78}
E Corrigan, D B Fairlie, P Goddard and S Templeton, A Green function
for the general self-dual gauge field. {\it Nucl Phys B} {\bf140} (1978) 31-44.
  
\bibitem{CG84}
E Corrigan and P Goddard, Construction of instanton and monopole solutions
and reciprocity. {\it Ann Phys} {\bf154} (1984) 253--279.

\bibitem{D92} S K Donaldson, Boundary value problems for Yang-Mills fields.
  {\it J Geom Phys} {\bf8} (1992) 89--122.
  
\bibitem{D10} S K Donaldson, Nahm's equations and free-boundary problems.
In: {\it The Many Facets of Geometry},
eds O~Garc\'ia-Prada, J~P~Bourguignon and S~Salamon (Oxford
University Press, Oxford, 2010).

\bibitem{CF77}
D B Fairlie and E Corrigan, Scalar field theory and exact solutions
to a classical SU(2) gauge theory.
{\it Phys Lett B} {\bf67} (1977) 69--71.

\bibitem{H12} D G Harland, The large $N$ limit of the Nahm transform.
  {\it Commun Math Phys} {\bf311} (2012) 689--712.

\bibitem{HPS12} D G Harland, S Palmer and C S\"amann, Magnetic domains.
  {\it JHEP} {\bf1210:167} (2012).
  
\bibitem{J04} M Jardim, A survey on Nahm transform.
  {\it J Geom Phys} {\bf52} (2004), 313--327.
  
\bibitem{Muk81} S Mukai, Duality between $D(X)$ and $D(\hat{X})$
  with its application to Picard sheaves.
  {\it Nagoya Math J} {\bf81} (1981), 153--175.  

\bibitem{N80} W Nahm, A simple formalism for the BPS monopole.
  {\it Phys Lett B} {\bf90} (1980) 413--414.

\bibitem{N83}
W Nahm, All self-dual multimonopoles for arbitrary gauge groups.
In: {\it Structural Elements in Particle Physics and Statistical Mechanics},
eds J~Honerkamp et al, NATO ASI Series B{\bf82} (1983), 301--310.

\bibitem{Os82} H Osborn, On the Atiyah-Drinfeld-Hitchin-Manin
  construction for self-dual gauge fields.
  {\it Commun Math Phys} {\bf86} (1982) 195--219.

\bibitem{W90} R S Ward, Linearization of the SU($\infty$) Nahm equations.
  {\it Phys Lett B} {\bf234} (1990) 81--84.  
  
\bibitem{Wit79}
E Witten, Some remarks on the recent twistor space constructions.
In: {\it Complex manifold techniques in theoretical physics},
eds D~E~Lerner and P~D~Sommers (Pitman, London, 1979).
  
\end{thebibliography}
\end{document}